Subgraph Classification, Clustering and Centrality for a Degree Asymmetric Twitter Based Graph Case Study: Suicidality

Keith Andrew, Eric Steinfelds
Department of Physics and Astronomy
Western Kentucky University
Bowling Green, KY 42104
Karla M. Andrew
CUSD 76, Oakwood, IL 61858
Kay Opalenik
Southern Connecticut State University
New Haven, CT 06515

Abstract
We present some initial results from a case study in social media data harvesting and visualization utilizing the tools and analytical features of NodeXL applied to a degree asymmetric vertex graph set. We consider twitter graphs harvested for topics related to suicidal ideation, suicide attempts, self-harm and bullycide. While the twitter-sphere only captures a small and age biased sample of communications it is a readily available public database for a wealth of rich topics yielding a large sample set. All these topics gave rise to highly asymmetric vertex degree graphs and all shared the same general topological features. We find a strong preference for in degree vertex information transfer with a 4:25 out degree to in degree vertex ratio with a power law distribution. Overall there is a low global clustering coefficient average of 0.038 and a graph clustering density of 0.00034 for Clauset-Newman-Moore grouping with a maximum geodesic distance of 6. Eigenvector centrality does not give any large central impact vertices and betweenness centrality shows many bridging vertices indicating a sparse community structure. Parts of speech sentiment scores show a strong asymmetry of predominant negative scores for almost all word and word pairs with salience greater than one. We used an Hoaxy analysis to check for deliberate misinformation on these topics by a Twitter-Bot.

I.     Introduction

The rapid growth of Twitter based communication has provided a means for analyzing information exchange between individuals on several topics across numerous platforms. Many of the information exchanges that take place on Twitter can be represented as a weighted directed graph[1] of tweets, retweets and mentions and often exhibits an asymmetry in vertex degree[2]. The global and local patterns of structure in the graphs are indicators of some of the characteristics of the group exchanging information[3]. As exemplified by the classification work of Smith, et. al. at the Social Media Research Foundation[4] there are at least six common network structures observed. These structures can be expressed as strongly divided or polarized, largely unified, fragmented, multi-clustered, predominately outward directed hub or as a predominately inward directed hub. A large graph can be made up of a distribution of these subsets and they can dynamically change in time[5,6] and can be modeled by systems of differential equations[7,8] and can form giant subgraphs[9,10,11], exhibit power law scaling behavior[12,13], demonstrate changes that are similar to phase changes[14], lead to viral events that are non-Bayesian[15], can topologically change in connectedness[16], have rapid information diffusion[17] that may percolate[18] or they can constitute a small world[19] graph. In the sense that there is information diffusion in a graph there are often different mechanisms at work for different sending protocols and structures, i.e. hashtags vs. tweets etc.[20]

Observing the classification and dynamics of such a graph for certain topics can lead to an engagement methodology based on threshold identifiers designed to indicate the development of action items, such as in sales or stock transfers. Here we are interested in applying these classification dynamics to a case study on the topic of suicidal ideation related to tweets referencing suicide attempts and self-harm.

Exploring social media and twitter conversations for parts of speech (POS) indicative of behavior change is an active research area especially in health-related subjects[21] and marketing[22]. The tracking of suicide risk factors,[23,24,25] bullying and cyber bullying[26], self-image[27], suicidal ideation[28,29] and indicators of self-harm[30] on social media has become a highly focused area of investigation[31]. Exploring preventive actions[32,33,34] and their overall efficacy is a current and growing area of research[35]. This has included self-presented[36] and highly public social media exchanges around suicidal[37] and self-harm[38] events including post analysis[39,40]. Efforts to study the complex patterns associated with twitter communications have often used word order and proximity[41] to machine learning[42,43] and AI strategies[44] to overcome the large data issues associated with analyzing social media streams. Often the POS in the tweet are analyzed regarding sentiment, syntax and salience within the context of a complex communication vertex graph or sociogram. The impact and influence of a vertex connected to certain POS can also be examined through frequency distributions and centrality measures. This coupled with knowledge about the local structure of the group can be used to help determine the nature of the interactions[45] and the diffusion rate[46] of information[47], which could be extensive if the graph is a governed by a power law[48,49] or is a small world graph[50]. Here we will examine the subgroup classification, vertex division and centrality measures in a case study of twitter networks and word searches related to suicide attempts and self-harm. We recognize that the twitter samples represent a biased subset of social media[51] users in the US: an estimated 7% compared to 41% on Facebook, and these communications are predominately used by younger affluent followers representing a distinct digital divide issue[52].

II.  SM Data Collection: NodeXL

Our data sets were collected using the social media data collection and analysis program NodeXL[53] from the Social Media Research Foundation.[54] The general features of this program allow users to map sociogram networks and provides tools to measure and understand the landscape of social media. NodeXL is an Excel based program designed to conduct social network analysis (SNA) including community clustering, influencer detection, content analysis, sentiment ranking and time series analysis. The sociogram networks can be displayed several ways including the Frutcherman-Reingold and the Harel-Koren Fast Multiscale vertex degree graphs with adjustable vertex forcing. The vertex graph can be analyzed for clustering effects and grouping which includes a Clauset-Newman-Moore, Wakita-Tsurumi or Girvan-Newman analysis or a clustering by a selected attribute of a vertex or edge. Data can be organized by tweets, mentions, retweets or hashtags with degree data stored for vertex or edge analysis. Sentiment, salience and centrality calculations can be carried out and used for sociogram plotting in terms of vertex size, color or shape allowing for higher dimensionality graphs. Tweets, Twitter links and image files are also collected during the tweet scan. Sentiment analysis can be carried out across any word set where we used the negative/positive, angry/calm and the twelve risk factor categories as search attributes for single words and word pairs with mutual influence. The Hoaxy[55] software routine was used to check very high tweet nodes for deliberate misinformation and/or manipulation of social media or Twitter-Bot detection for the sources related to the self-harm and suicide-attempts topics.

III.   Clustering, Classification and Centrality Analytics of Sociograms

**Clustering**
Tweets were gathered starting 5/27/2015 for searches related to tweets including words such as: suicide, suicide attempts, bulleycide, drug suicide/abuse, suicidal thoughts, self-cutting, self-pain, veterans suicide, and survivor topics.  The initial graphs of V vertices and E edges, G(V, E) for suicide attempts were collected as sociograms using a Harel-Koren Fast Multiscale visualization and with a Frutcherman-Reingold graph with a modified repulsive index of 3.5 after ten iterations, are shown in Fig. (1). Each vertex represents a user and each edge corresponds to a tweet, mention, replies-to or retweet with the arrow indicating the vertex of origin to the receiving vertex.  There are 2494 vertices with 2602 edges of which 170 are duplicates giving 2432 unique edges in Fig. (1) (A).  The groups in Fig. (1) were determined by the Clauset-Newman-Moore algorithm[56] using decision rules based on edge density modularity[57] to determine community structures or clusters where each group is visualized in a separate box and labeled with a common vertex word from the corresponding tweets between members. Fig. (1) is grouped by top words in tweets listed by salience.  The corresponding cluster coefficient is defined as the ratio of closed vertex triplets to all possible triplets, open or closed:

$$\text{at each vertex on a graph G(V,E)}: c_j = \frac{n_j(closed-triplets)}{n(all-triplets)}$$

$$\text{Global Clustering Coefficient:Average:} C_{global} = \sum_{j=1}^{N} \frac{c_j}{N_{vertices}}$$

$$\text{Directed Graph Density}: \rho(E,V) = \frac{|E|}{|V|(|V|-1)}$$

$$\text{Maximum number of edges}: N_{Edges} = \frac{|V|(|V|-1)}{2}$$

(1)

For the graph in Fig. (1) (A) the global average clustering coefficient is $C_{global}=0.038$ with a graph density of $\rho = 0.000346$, an average geodesic distance of $d_{avg} = 1.989$ and a maximum geodesic distance or graph diameter of $D = 6$. For a graph of 2494 vertices this indicates a weakly clustered not strongly connected set.  Most vertices have a small number of one-way connections without large numbers of growing retweets, followers or mentions. This global structure was similar for each related search topic with the largest well-connected subgroup almost always appearing as the group of external health care providers and various survivor networks.

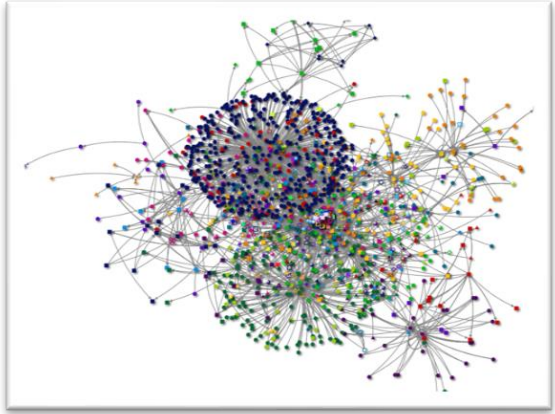 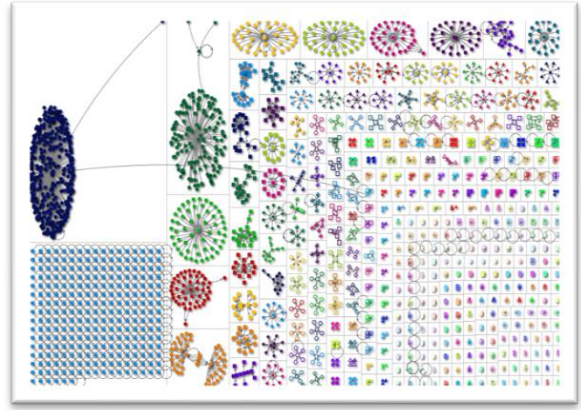

(A) Directed FR: 2494 vertices, 2602 edges    (B) Directed FR 831 subgroups by salience

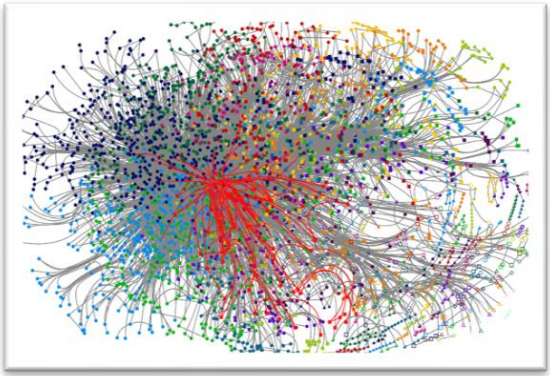 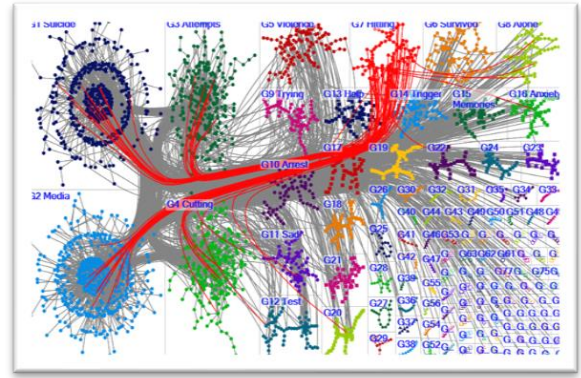

(C) Directed HK:3,336 vertices 11,138 edges    (D) HK 123 Groups with key word labels

Fig. (1) Left: (A) A directed graph Sociogram for Twitter Search Network key: "Suicide Attempts" displayed in Frutcherman-Reingold format for 2494 vertices from NodeXL, Right (B) the cluster decomposition to subgroups based on common topic. (C) the open sociogram on self-harm with connections to hitting highlighted in Harel-Koren form and (D) the group structure showing the edges that link self-harm to other groups with the same key words.

Each global graph can be broken down into clusters and subgraphs based upon an ordering attribute to assist in analyzing the community structure and interactions within the graph. The six graph types appear with a varying frequency as subgraphs depending upon the type of interactions and communities the individuals are involved in. For the combined global graphs: suicide attempts, self-hurt, bulleycide, veterans, teen suicide and suicide we find an overall distribution given in Fig. (2). Fig.(2) (A) displays the six common types of graphs: a bounded unified graph, a hub centered graph with inward directed spokes, a hub centered graph with outward directed spokes, a multitopic graph with several distinct active vertices and some isolated outliers, a polarized graph with two distinct but opposed groups with few cross pole interactions, and a fragmented group graph with many small isolated vertices often with numerous self-tweets. This pattern remained unchanged during the data collection period from 2015 until Dec. 2019 for each set of graphs collected. A summary of the main features of the vertex graphs for the six search topics: Suicide Attempt, Suicide Self Harm, Suicidal Ideation, Suicide Drugs, Suicide Firearms and Suicide Violence are given in Table 1.

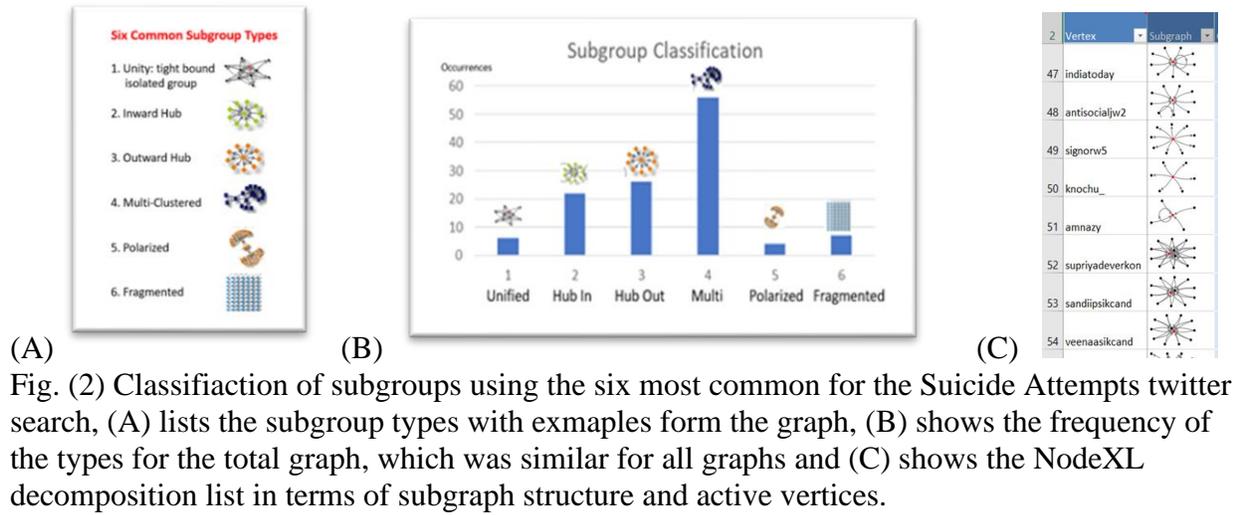

(A)  (B)  (C)

Fig. (2) Classifiaction of subgroups using the six most common for the Suicide Attempts twitter search, (A) lists the subgroup types with exmaples form the graph, (B) shows the frequency of the types for the total graph, which was similar for all graphs and (C) shows the NodeXL decomposition list in terms of subgraph structure and active vertices.

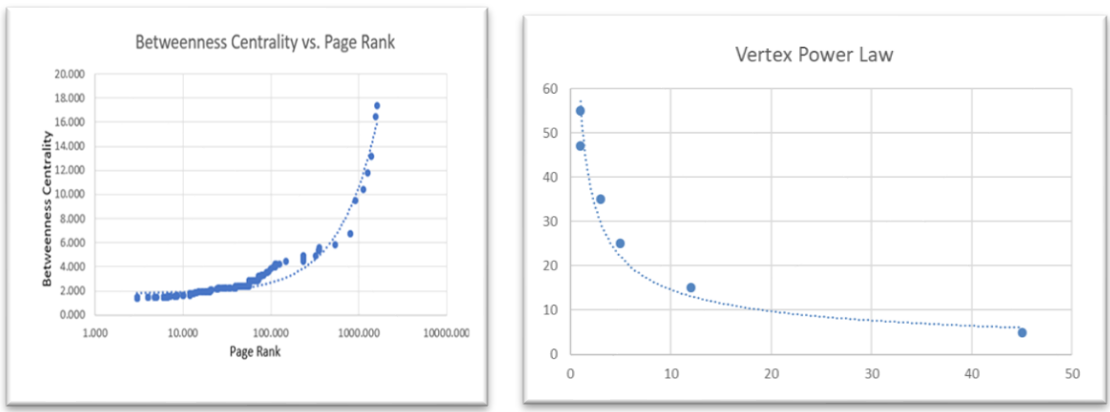

Fig. (3) Both the Page Rank as a function of Betweenness Centrality and the vertex degree obey power law distributions.

|   | Suicide Attempt | Suicide Self Harm | Suicidal Ideation | Suicide Drugs | Suicide Firearms | Suicide Violence |
|---|---|---|---|---|---|---|
| V | 2494 | 2497 | 2275 | 1420 | 2162 | 2611 |
| E | 2432 | 2283 | 2168 | 1551 | 2015 | 2209 |
| SG | 831 | 414 | 701 | 191 | 319 | 895 |
| C | 0.038 | 0.025 | 0.009 | 0.041 | 0.003 | 0.017 |
| D | 6 | 8 | 4 | 11 | 18 | 10 |
| ρ | 0.000345 | 0.0003034 | 0.000328 | 0.000718 | 0.00041 | 0.00027 |

Table 1 Summary of Graph features for six Twitter based search topics.

Total tweet level activity can be characterized by the relative In and Out degrees of the vertices and by various centrality measures of the graph. As indicated by Fig. (3) for these specific graphs the betweenness centrality grows as a power law with page rank, which is the principle eigenvector of the normalized adjacency matrix for the graph and where vertex betweenness is defined as

$$Betweenness: C_B(v) = \sum_{s \neq t \neq v} \frac{n_{st}(v)}{n_{st}}$$

$n_{st}$ = total number of shortest paths from s to t

$n_{st}(v)$ = number of those paths through node v  (2)

from graph G(V,E) in Fig. (1) (C) we have for page rank $r_p$: $C_B = ke^{\alpha r_p}$

$where: k = 1.89 \quad \alpha = 0.002 \quad R^2 = 0.67$ .

The vertex degree distribution power law describes the global graph behavior. Here we have a power law where

$$Vertex\ Power\ Law: P(x) = ax^{-\beta x}$$

Total Vertex Degree: $a = 57.1 \quad \beta = 0.59 \quad R^2 = 0.96$ . (3)

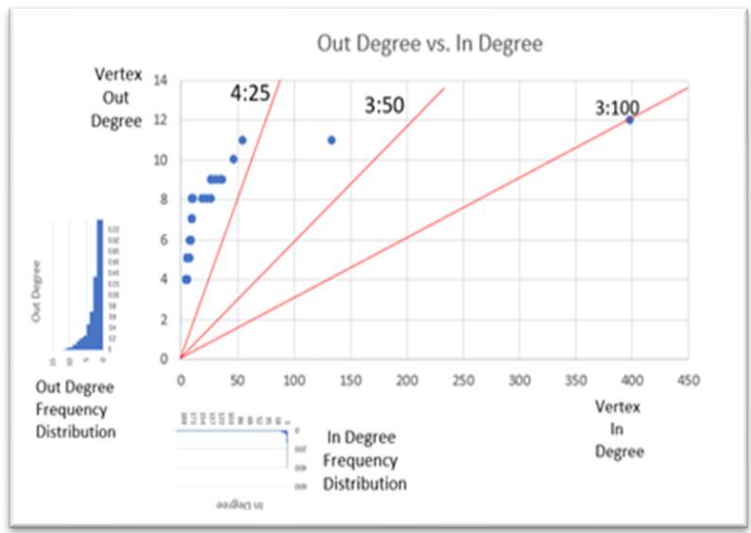

Fig. (4) Vertex Degree Separation of users into groups of followers, In-degree, and those who are following, Out-degree, with regions corresponding to the ratio of Out/In, or dy/dx, where the axes are scaled to show the difference and with insets for the frequency distribution of Out-degree and In-degree. In-degree vertices dominate all of our graphs on these topics (we have removed the cluster of points near the origin for clarity).

**Vertex Degree**
The values of Fig. (4) demonstrate the asymmetric nature of the activity of the vertices in the Suicide Attempts and Self Harm Sociograms. The vast majority of the vertices are sending messages in, typically six times more than they have tweets going out. Overall the groups are relatively small with the largest subgraph being from the collective of suicide prevention providers. The twitter activity of a given user can be monitored outside of this narrow topic and not surprisingly some are quite prolific compared to their activity in this graph. From the In-degree inset plot in Fig. (4) we see that the frequency drops off very quickly when compared to the Out-degree frequency. This asymmetry manifests itself in the Out-degree vs. In-degree plot by placing most users in the 4:25 region indicating four tweets sent out for each 25 received, or four users being followed for each 25 following. The vertex asymmetry is measured by

$$\text{vertex asymmetry strength } R_v = \log\left(\frac{n_{in}}{n_{out}}\right) \quad (4)$$

|  | Attempt | Self Harm | Ideation | Drugs | Firearms | Violence |
|---|---|---|---|---|---|---|
| In | 2498 | 2273 | 2189 | 1983 | 3023 | 2317 |
| Out | 1910 | 1358 | 1071 | 726 | 1648 | 1105 |
| $R_{vertex}$ | 0.117 | 0.224 | 0.310 | 0.436 | 0.263 | 0.322 |

Table 2 Global Vertex Asymmetry Strengths

This In-degree and Out-degree asymmetry is also evident in Fig. (5) where the vertex degree is indicated by the size of the vertex plotted in the graph. The sociogram for In-degree displays many medium size vertices and only a few large central ones. In the Out-degree sociogram there are a few large vertices followed by several medium vertices spread throughout the graph followed by numerous small vertices. This pattern again emphasizes the 4:25 tweet differential of four following for each 25 followers for most vertices.

**Centrality**
Fig. (5) also shows the centrality measures for the total graph. For betweenness centrality the size of the vertex indicates the magnitude of the betweenness centrality. For these topics there are a number of vertices with high betweenness centrality indicating the noncentralized nature of the graph and users. These vertices serve as transition points from one topical cluster to another in the graph. These vertices match the highly fragmented nature of this graph compared to other project topics such as the Banned Books and Arab Spring graphs which exhibit lower betweenness and nearly equal in and out degrees. The eigenvector centrality graph shows a large number of medium and small vertices, many of which are clumped in an inward hub giant subgraph. The vertices with the highest eigenvector centrality are the ones with the largest connectivity and highest impact on neighboring vertices. Here the eigenvector centrality shows the fragmented nature of these weakly interacting communities linked by the tweeting on suicidality and self-harm but with no subset of very strong centralized users.

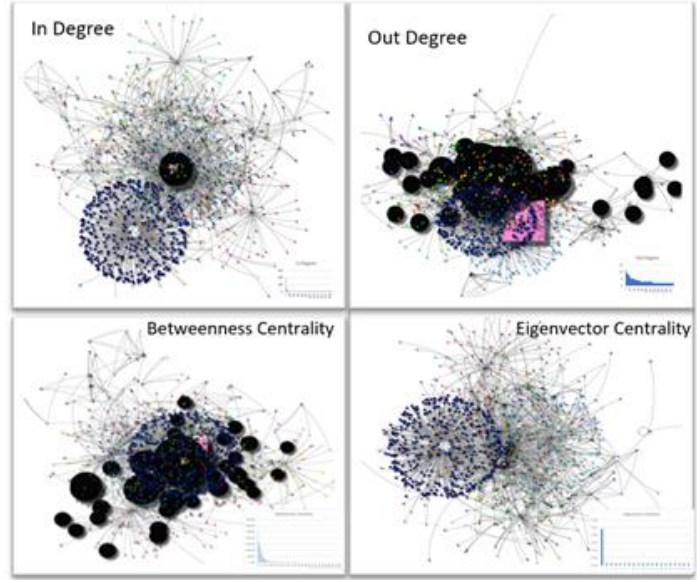

Fig. (5) Activity Profiles with distribution insets. In each graph the vertex size indicates the value of the listed attribute for degree in top row and for centrality in bottom row.

**Parts of Speech**
To help understand the topics within the groups we examined the word frequency, salience and sentiment of the most frequent words and word pairs or bigrams. Sentiment ratios of these graphs are compared to earlier studies on Banned Books and the Arab Spring twitter-based graphs.

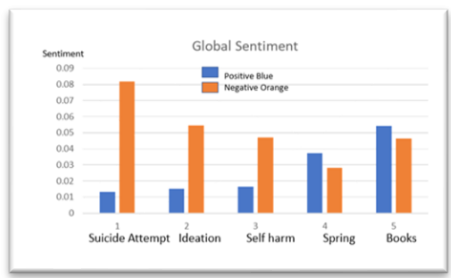

| Global Sentiment | Value |
|---|---|
| Positive | 0.012 |
| Negative | 0.061 |
| Anger Mild | 0.011 |
| Anger Typical | 0.029 |
| Anger Violent | 0.027 |

(A)      (B)      (C)

| Cluster Group: type | Word Pair | Salience | Mutual Information |
|---|---|---|---|
| G2: out hub | suicide attempt | 0.011 | 1.101 |
| G6: multi | repeatedly mocked | 0.015 | 1.255 |
| G3: multi | vulnerable person | 0.015 | 1.255 |
| G8: fragmented | daughter's suicide | 0.016 | 1.431 |
| G4: multi | depressed life | 0.015 | 1.161 |
| G12: multi | uncontrollable sadness | 0.014 | 1.301 |
| G15: multi | reach out | 0.013 | 1.342 |
| G21: multi | suicide note | 0.010 | 1.130 |
| G9: multi | emotionally abusive | 0.011 | 1.121 |
| G3: multi | violent outbursts | 0.012 | 1.262 |

Fig (6) Global sentiment measures for positive and negative scores and for level of anger separated into low level or mild anger, mid-level typical anger and high-level violent anger for Suicide Attempts graph in (A), (B) shows the sentiment scores for three recent graphs: Suicide Attempts, Suicidal Ideation and Self-Harm compared to a graph on Arab Spring and one on Banned Books, (C) shows the bigram word pairs with their salience and associated mutual information for the average of all three graphs- Attempts, Ideation and Harm.

The sentiment strength asymmetry ratios for positive/negative and anger/calm sentiments are given by

$$R_{sent} = \log_{10}\left(\frac{s_{pos}}{s_{neg}}\right) \quad R_{angr} = \log_{10}\left(\frac{s_{calm}}{s_{angr}}\right) \quad (5)$$

|  | Attempt | Self Harm | Ideation | Drugs | Firearms | Violence |
|---|---|---|---|---|---|---|
| $R_{sent}$ | -0.71 | -0.75 | -0.84 | -0.63 | -0.68 | -0.72 |
| $R_{anger}$ | -0.82 | -0.72 | -0.72 | -0.71 | -0.64 | -0.75 |

Table 3. Global strength ratios for positive/negative and anger/calm sentiment values for each graph.

While word sentiment is an effort to measure some of the emotional direction and level of words in a sentence structure compared to a given list- i.e. negative/positive or anger/calm, salience is a measure of importance or impact based upon the relative frequency of use. When used to analyze tweets the overall sentence structure and context is often cryptic and hidden in many symbols and/or emoticons. Many words will have more significance when paired in specific two word phrases, captured in part by the two word mutual information factor, which is higher for pairs that have a stronger and more impactful presence in a tweet. For all of the graphs we examined the word pair: Suicide Attempt, has a higher mutual information rating than each word alone. Several common phases that appeared repeatedly are listed in Fig. (6) (C).

|  | Risk Factors | Word Identifiers | Tweets | Salience |
|---|---|---|---|---|
| 1. | Depressive feelings | Alone, depressed, helpless, empty, sad, anxiety, apathy, restless, agitation | 34 | 0.001 |
| 2. | Depressive symptoms | Sleeping a lot, irritable, tired, , agitation, crying, restless, insomnia | 1884 | 0.045 |
| 3. | Drug abuse | Depressed, alcohol, sertraline, Zoloft, Prozac, pills, drugs, meth, coke, high | 24 | 0.001 |
| 4. | Prior suicide attempt | Suicide again, attempt, try, commit | 2041 | 0.049 |
| 5. | Suicide around individual | Mom, mother, brother, sister, dad, father, friend, uncle, know someone | 117 | 0.003 |
| 6. | Ideation | Thought of suicide, thoughts of killing myself, want to commit, thinking | 2695 | 0.064 |
| 7. | Self harm | Stop cutting myself, cutting, slit, hair pull, injure, harm, nails, burn, strain, break, bruise, disable, bones, | 78 | 0.001 |
| 8. | Bullying | Being bullied, feel bullied, stop bullying, getting bullied, picked on, | 409 | 0.010 |
| 9. | Gun ownership | Gun suicide, bullets, shot, shoot, wound, | 1180 | 0.028 |
| 10. | Psychological disorder | Diagnosed, anorexia, bipolar, ocd, weight, adhd, | 37 | 0.001 |
| 11. | Family violence/ discord | Parents fight, fighting, boy/girl friend fight, black eye, argue | 65 | 0.002 |
| 12. | Impulsive | I am impulsive, hasty, passionate, emotional, impetuous, suddenly | 1076 | 0.026 |

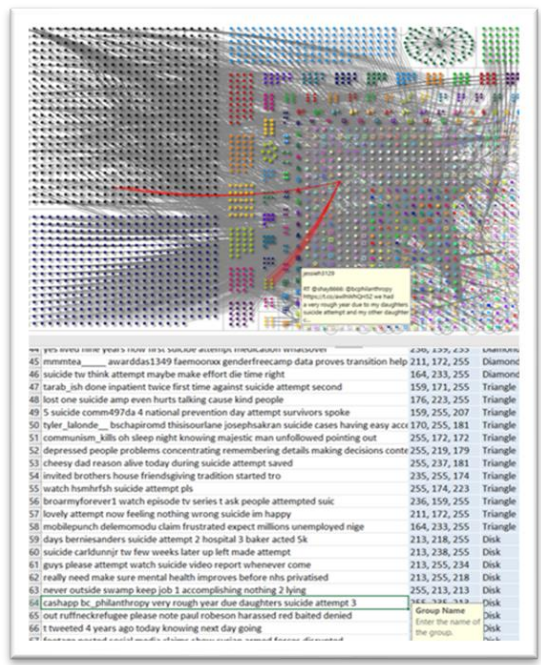

Fig. (7) The dozen risk factors from Jashinsky appearing in the graph as a function of identified tweets and overall salience. For the graph representation in terms of vertex attribute for salience each tweet/retweet bundle links to a common group, On the right the vertex for a mother seeing her daughter in a suicide attempt links directly to the retweet groups that formed.

Coupling the risk factors to the eigenvector centrality and page rank of a vertex in a subgroup gives an ordering of groups indicative of the highest risk factor group of tweets, retweets and mentions for the entire graph. As indicated in Fig. (7), in terms of single and double word searches there are five risk factors which appear with high frequency in every search topic we examined: depressive symptoms, prior suicide attempts, suicidal ideation, gun access, and impulsiveness. For the graphs from Suicidality, Suicide Attempts, Self-Harm and Bullycide some 58% of the subgroup community clusters explicitly tweet about three or more of the five highest ranked risk factors using terms listed in Fig. (7). These more active groups provide a subset of users whose tweets can be more carefully analyzed and studied by both tracking the tweet dynamics and visiting many of the associated Twitter sites.

To check our most used twitter feeds for misinformation and bot-like behavior we performed corresponding searches using the Hoaxy analysis package. The sites that were flagged as being potential problems, seen as red vertices in Fig. (8) were almost all inactive feeds with no followers, mentions or retweets. There were three sites related to strong media and blogging activity on suicide, we could not find a vertex associated with a large number of potentially automated suicide tweets.

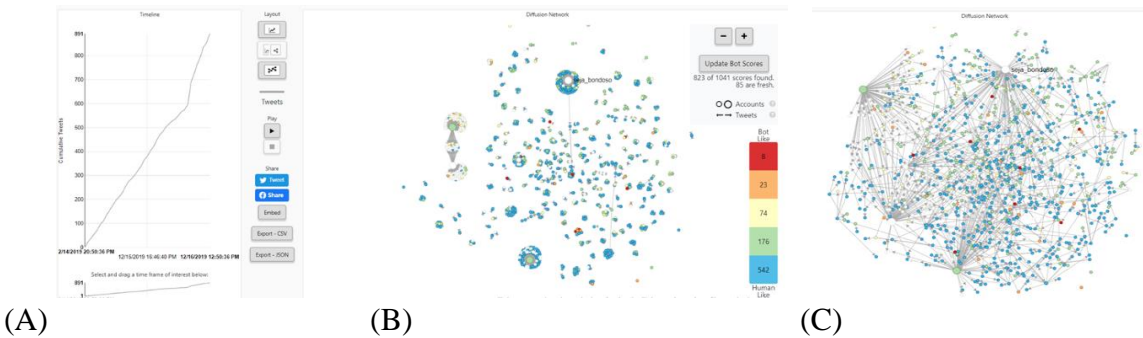

(A)          (B)          (C)

Fig. (8) Hoaxy Diffusion Plots and Timeline for Suicide Attempts twitter search. (A) Is a representative plot of the cumulative number of tweets related to Suicide Attempts starting at 20:50:36 CST on 12/14/2019 and ending 1t 12:50:36 on 12/16/2019 reaching a maximum of 891. (B) and (C) shows the resulting fractured diffusion network at different scales indicating one polarized cluster and two isolated hub clusters with four potential Twitter-Bot vertices.

## IV.    Conclusions

We have examined vertex graph sociogram characteristics for twitter users interacting on topics from suicidal ideation and self-harm as a case study on asymmetric graph features using NodeXL. For each sociogram produced in each topic area: Suicidality, Suicide Attempts, Self-Harm and Bullycide we find a fragmented weakly connected set of communities with vertex degree asymmetry favoring in-degree, low overall eigenvector centrality, low density and low clustering coefficients. For high salience words and bigrams the graphs exhibit a sentiment asymmetry that is also exhibited at the subgraph level. The highest salience words closely match the risk indicators of Jashinksy and multiple repeats from four or more categories for subgraphs is not unusual. These subgroups form an especially interesting and active set of tweeters for these topics and provide an excellent group for closer study.

V.　　References

[1] Cogan, Peter, Matthew Andrews, Milan Bradonjic, W. Sean Kennedy, Alessandra Sala, and Gabriel Tucci. "Reconstruction and analysis of twitter conversation graphs." In *Proceedings of the First ACM International Workshop on Hot Topics on Interdisciplinary Social Networks Research*, pp. 25-31. ACM, 2012.

[2] Erdős, Paul, and Alfréd Rényi. "Asymmetric graphs." *Acta Mathematica Hungarica* 14, no. 3-4 (1963): 295-315.

[3] Ardon, Sebastien, Amitabha Bagchi, Anirban Mahanti, Amit Ruhela, Aaditeshwar Seth, Rudra Mohan Tripathy, and Sipat Triukose. "Spatio-temporal and events based analysis of topic popularity in twitter." In *Proceedings of the 22nd ACM international conference on Information & Knowledge Management*, pp. 219-228. ACM, 2013.

[4] Smith, Marc A., Lee Rainie, Ben Shneiderman, and Itai Himelboim. "Mapping Twitter topic networks: From polarized crowds to community clusters." *Pew Research Center* 20 (2014): 1-56.

[5] Amati, Giambattista, Simone Angelini, Francesca Capri, Giorgio Gambosi, Gianluca Rossi, and Paola Vocca. "TWITTER TEMPORAL EVOLUTION ANALYSIS: COMPARING EVENT AND TOPIC DRIVEN RETWEET GRAPHS." *IADIS International Journal on Computer Science & Information Systems* 11, no. 2 (2016).

[6] Smailhodvic, Armin, Keith Andrew, Lance Hahn, Phillip C. Womble, and Cathleen Webb. "Sample NLPDE and NLODE Social-Media Modeling of Information Transmission for Infectious Diseases: Case Study Ebola." *arXiv preprint arXiv:1501.00198* (2014).

[7] Wang, Haiyan, Feng Wang, and Kuai Xu. "Modeling information diffusion in online social networks with partial differential equations." *arXiv preprint arXiv:1310.0505* (2013).

[8] Taylor, Morgan, Armin Smailhodzic, Keith Andrew, Lance Hahn, Phil Womble, Cathleen Webb, and Blair Thompson. "A System of coupled ODEs as a Cyber Model for Analyzing Wavelike Information Transmission from Data Mining Tweets." *Bulletin of the American Physical Society* 59 (2014).

[9] Frieze, Alan, Michael Krivelevich, and Ryan Martin. "The emergence of a giant component in random subgraphs of pseudo-random graphs." *Random Structures & Algorithms* 24, no. 1 (2004): 42-50.

[10] Chung, Fan, Paul Horn, and Linyuan Lu. "The giant component in a random subgraph of a given graph." In *International Workshop on Algorithms and Models for the Web-Graph*, pp. 38-49. Springer, Berlin, Heidelberg, 2009.

[11] Andrew, Keith, Morgan Taylor, Phillip Womble, Karla Andrew, Kay Opalenik, and Craig Cobane. "Percolation Threshold from a Giant Subgraph of a Twitter Based Nodal Graph." *Bulletin of the American Physical Society* 62 (2017).

[12] Bild, David R., Yue Liu, Robert P. Dick, Z. Morley Mao, and Dan S. Wallach. "Aggregate characterization of user behavior in Twitter and analysis of the retweet graph." *ACM Transactions on Internet Technology (TOIT)* 15, no. 1 (2015): 4.

[13] Gonzalez, Joseph E., Yucheng Low, Haijie Gu, Danny Bickson, and Carlos Guestrin. "Powergraph: Distributed graph-parallel computation on natural graphs." In *Presented as part of the 10th {USENIX} Symposium on Operating Systems Design and Implementation ({OSDI} 12)*, pp. 17-30. 2012.

[14] Andrew, Keith, Morgan Taylor, Karla Andrew, and Phillip Womble. "Diffusive Phase Change Model of Twitter Information in the Context of a Cusp Catastrophe." *Bulletin of the American Physical Society* 61 (2016).

[15] Hansen, Lars Kai, Adam Arvidsson, Finn Årup Nielsen, Elanor Colleoni, and Michael Etter. "Good friends, bad news-affect and virality in twitter." In *Future information technology*, pp. 34-43. Springer, Berlin, Heidelberg, 2011.

[16] Antoniades, Demetris, and Constantine Dovrolis. "Co-evolutionary dynamics in social networks: A case study of twitter." *Computational Social Networks* 2, no. 1 (2015): 14.

[17] Smailhodzic, Armin, Keith Andrew, Eric Steinfelds, Lance Hahn, Phil Womble, and Cathleen Webb. "A Wave Equation Cyber Model for Tracking Ebola from Data Mining African Tweets." *Bulletin of the American Physical Society* 59 (2014).

[18] Piraveenan, Mahendra, Mikhail Prokopenko, and Liaquat Hossain. "Percolation centrality: Quantifying graph-theoretic impact of nodes during percolation in networks." *PloS one* 8, no. 1 (2013): e53095.

[19] Ye, Shaozhi, and S. Felix Wu. "Measuring message propagation and social influence on Twitter. com." In *International conference on social informatics*, pp. 216-231. Springer, Berlin, Heidelberg, 2010.

[20] Romero, Daniel M., Brendan Meeder, and Jon Kleinberg. "Differences in the mechanics of information diffusion across topics: idioms, political hashtags, and complex contagion on twitter." In *Proceedings of the 20th international conference on World wide web*, pp. 695-704. ACM, 2011.

[21] Korda, Holly, and Zena Itani. "Harnessing social media for health promotion and behavior change." *Health promotion practice* 14, no. 1 (2013): 15-23.

[22] Goh, Khim-Yong, Cheng-Suang Heng, and Zhijie Lin. "Social media brand community and consumer behavior: Quantifying the relative impact of user-and marketer-generated content." *Information Systems Research* 24, no. 1 (2013): 88-107.

[23] Jashinsky, Jared, Scott H. Burton, Carl L. Hanson, Josh West, Christophe Giraud-Carrier, Michael D. Barnes, and Trenton Argyle. "Tracking suicide risk factors through Twitter in the US." *Crisis* (2014).

[24] Fodeh, Samah, Joseph Goulet, Cynthia Brandt, and Al-Talib Hamada. "Leveraging Twitter to better identify suicide risk." In *Medical Informatics and Healthcare*, pp. 1-7. 2017.

[25] Du, Jingcheng, Yaoyun Zhang, Jianhong Luo, Yuxi Jia, Qiang Wei, Cui Tao, and Hua Xu. "Extracting psychiatric stressors for suicide from social media using deep learning." *BMC medical informatics and decision making* 18, no. 2 (2018): 43.

[26] Litwiller, Brett J., and Amy M. Brausch. "Cyber bullying and physical bullying in adolescent suicide: the role of violent behavior and substance use." *Journal of youth and adolescence* 42, no. 5 (2013): 675-684.

[27] Brausch, Amy M., and Jennifer J. Muehlenkamp. "Body image and suicidal ideation in adolescents." *Body image* 4, no. 2 (2007): 207-212.

[28] De Choudhury, Munmun, Emre Kiciman, Mark Dredze, Glen Coppersmith, and Mrinal Kumar. "Discovering shifts to suicidal ideation from mental health content in social media."

In *Proceedings of the 2016 CHI conference on human factors in computing systems*, pp. 2098-2110. ACM, 2016.

[29] O'Dea, Bridianne, Stephen Wan, Philip J. Batterham, Alison L. Calear, Cecile Paris, and Helen Christensen. "Detecting suicidality on Twitter." *Internet Interventions* 2, no. 2 (2015): 183-188.

[30] Emma Hilton, Charlotte. "Unveiling self-harm behaviour: what can social media site Twitter tell us about self-harm? A qualitative exploration." *Journal of clinical nursing* 26, no. 11-12 (2017): 1690-1704.

[31] Luxton, David D., Jennifer D. June, and Jonathan M. Fairall. "Social media and suicide: a public health perspective." *American journal of public health* 102, no. S2 (2012): S195-S200.

[32] Robinson, Jo, Georgina Cox, Eleanor Bailey, Sarah Hetrick, Maria Rodrigues, Steve Fisher, and Helen Herrman. "Social media and suicide prevention: a systematic review." *Early intervention in psychiatry* 10, no. 2 (2016): 103-121.

[33] Abboute, Amayas, Yasser Boudjeriou, Gilles Entringer, Jérôme Azé, Sandra Bringay, and Pascal Poncelet. "Mining twitter for suicide prevention." In *International Conference on Applications of Natural Language to Data Bases/Information Systems*, pp. 250-253. Springer, Cham, 2014.

[34] Fodeh, Samah Jamal, Edwin D. Boudreaux, Rixin Wang, Dennis Silva, Robert Bossarte, Joseph Lucien Goulet, Cynthia Brandt, and Hamada Hamid Altalib. "Suicide Risk on Twitter." *International Journal of Knowledge Discovery in Bioinformatics (IJKDB)* 8, no. 2 (2018): 1-17.

[35] Varathan, Kasturi Dewi, and Nurhafizah Talib. "Suicide detection system based on Twitter." In *2014 Science and Information Conference*, pp. 785-788. IEEE, 2014.

[36] Fu, King-wa, Qijin Cheng, Paul WC Wong, and Paul SF Yip. "Responses to a self-presented suicide attempt in social media." *Crisis* (2013).

[37] Ueda, Michiko, Kota Mori, Tetsuya Matsubayashi, and Yasuyuki Sawada. "Tweeting celebrity suicides: Users' reaction to prominent suicide deaths on Twitter and subsequent increases in actual suicides." *Social Science & Medicine* 189 (2017): 158-166.

[38] Baker, Darren, and Sarah Fortune. "Understanding self-harm and suicide websites: A qualitative interview study of young adult website users." *Crisis* 29, no. 3 (2008): 118-122.

[39] Gunn, John F., and David Lester. "Twitter postings and suicide: An analysis of the postings of a fatal suicide in the 24 hours prior to death." *Suicidologi* 17, no. 3 (2015).

[40] Sueki, Hajime. "The association of suicide-related Twitter use with suicidal behaviour: a cross-sectional study of young internet users in Japan." *Journal of affective disorders* 170 (2015): 155-160.

[41] Hahn, Lance W., and Robert M. Sivley. "Entropy, semantic relatedness and proximity." *Behavior research methods* 43, no. 3 (2011): 746-760.

[42] Braithwaite, Scott R., Christophe Giraud-Carrier, Josh West, Michael D. Barnes, and Carl Lee Hanson. "Validating machine learning algorithms for Twitter data against established measures of suicidality." *JMIR mental health* 3, no. 2 (2016): e21.

[43] Burnap, Pete, Walter Colombo, and Jonathan Scourfield. "Machine classification and analysis of suicide-related communication on twitter." In *Proceedings of the 26th ACM conference on hypertext & social media*, pp. 75-84. ACM, 2015.

[44] Fonseka, Trehani M., Venkat Bhat, and Sidney H. Kennedy. "The utility of artificial intelligence in suicide risk prediction and the management of suicidal behaviors." *Australian & New Zealand Journal of Psychiatry* 53, no. 10 (2019): 954-964.

[45] Ediger, David, Karl Jiang, Jason Riedy, David A. Bader, Courtney Corley, Rob Farber, and William N. Reynolds. "Massive social network analysis: Mining twitter for social good." In *2010 39th International Conference on Parallel Processing*, pp. 583-593. IEEE, 2010.

[46] Yang, Jiang, and Scott Counts. "Predicting the speed, scale, and range of information diffusion in twitter." In *Fourth International AAAI Conference on Weblogs and Social Media*. 2010.

[47] Kawamoto, Tatsuro. "A stochastic model of tweet diffusion on the Twitter network." *Physica A: Statistical Mechanics and its Applications* 392, no. 16 (2013): 3470-3475.

[48] Otte, Evelien, and Ronald Rousseau. "Social network analysis: a powerful strategy, also for the information sciences." *Journal of information Science* 28, no. 6 (2002): 441-453.

[49] Milojević, Staša. "Power law distributions in information science: Making the case for logarithmic binning." *Journal of the American Society for Information Science and Technology* 61, no. 12 (2010): 2417-2425.

[50] Watts, Duncan J., and Steven H. Strogatz. "Collective dynamics of 'small-world' networks." *nature* 393, no. 6684 (1998): 440.

[51] Twitter by the Numbers, https://www.omnicoreagency.com/twitter-statistics/

[52] Blank, Grant. "The digital divide among Twitter users and its implications for social research." *Social Science Computer Review* 35, no. 6 (2017): 679-697.

[53] NodeXL at https://nodexl.com/

[54] Social Media Research Foundation at https://www.smrfoundation.org/

[55] Observatory on Social Media, OSoMe at the Network Science Institute at IUNI, the Center for complex Networks and Systems Research, CNetS at SICE and the Media School at Indiana University: https://hoaxy.iuni.iu.edu/

[56] Clauset, Aaron, Mark EJ Newman, and Cristopher Moore. "Finding community structure in very large networks." *Physical review E* 70, no. 6 (2004): 066111.

[57] Stanford Network Analysis Platform, http://snap.stanford.edu/